\documentclass[pra,twocolumn,showpacs]{revtex4}

\usepackage{graphicx}
\usepackage{amsmath}

\begin{document}

\title{On the transverse mode of an atom laser}

\author{Th.~Busch$^{1[\dagger]}$, 
        M.~K\"ohl$^{2,3}$, 
        T.~Esslinger$^{2,3}$, and  
        K.~M\o lmer$^1$}

\affiliation{$^1$Institute of Physics and Astronomy, Aarhus University, 
                 Ny Munkegade, DK--8000 \AA rhus C, Denmark\\
             $^2$Sektion Physik, Ludwig--Maximilians--Universit\"at, 
                 Schellingstr.~4/III, D--80799 M\"unchen, Germany\\
             $^3$Institute of Quantum Electronics, ETH H\"onggerberg, 
                 CH--8093 Z\"urich, Switzerland}

\date{\today}

\begin{abstract}
  The transverse mode of an atom laser beam that is outcoupled from a
  Bose--Einstein condensate is investigated and is found to be
  strongly determined by the mean--field interaction of the laser beam
  with the condensate.  Since for repulsive interactions the geometry
  of the coupling scheme resembles an interferometer in momentum
  space, the beam is found show filamentation.  Observation of this
  effect would prove the transverse coherence of an atom laser beam.
\end{abstract}

\pacs{03.75.Fi, 32.80.-t}

\maketitle

Research on atom lasers is an active and fascinating area in atomic
physics \cite{Ketterle:97,Esslinger:99}.  Several laboratories around
the world are now using continuous output couplers to produce atom
laser beams from Bose-Einstein condensates.  It is therefore important
to characterize the qualities of these beams.  Recently, their
temporal coherence was verified \cite{Esslinger:01} and their
transverse divergence was measured \cite{Aspect:01}.

Although atom lasers and optical lasers show strong similarities, the
possibility for atoms to scatter off each other leads to various
effects absent in optical lasers. In the following we will show that a
strong, inhomogeneous repulsive potential, as is represented by the
remaining Bose-Einstein condensate, can be a source of instabilities
for the beam and, in particular, can lead to its transverse
filamentation. Since most experimental setups involve asymmetric traps
with two stiff directions, our results are immediately applicable to
these experiments.

In magnetic traps output couplers for atom laser beams are realized by
coupling a fraction of the Bose-Einstein condensate into a
magnetically untrapped state. This process happens inside the trapped
sample along a surface where the resonance condition for output
coupling is fulfilled, subjecting the output coupled atoms to the
repulsive mean field potential of the condensate. Since gravity
displaces the symmetry axis of the Bose-Einstein condensate with
respect to the symmetry axis of the magnetic trapping field, the
repulsive potential is however not homogeneous over the resonance
surface. In a classical picture this situation corresponds to
particles rolling off a potential from different heights, leading to a
non--negligible momentum spread or dispersion in the transverse
directions \cite{MomentumSpread}. Moreover, the \textsl{finite}
interaction time of the falling beam with the remaining condensate
leads to a non--monotonic increase of the transverse atomic
position $x(t)$ as a function of the points of resonant output
coupling $x_i$. In the quantum dynamics, this leads to the
interference of atoms with different transverse momenta within the
beam.

In the following we will investigate this interference process. First
we consider an idealized, one--dimensional model in the direction
perpendicular to gravity, and we examine its classical,
semiclassical and quantum behaviour. After we have identified and
described the relevant processes, we will include the effects of
gravity and propagation in more than one dimension.

Atom lasers with radiofrequency output couplers usually couple
different Zeeman substates of the trapped atoms. For a Bose-Einstein
condensate in a $|F=1,m_F=-1\rangle$ state, two output states are
possible.  Either $m_F=0$ or $m_F=1$, the first of which has a
vanishing magnetic moment and the latter experiences a repulsive force
by the magnetic trap.  However, since the output coupling rate from
the $m_F=-1$ state into the $m_F=0$ state is usually chosen to be
small, subsequent transitions into the $m_F=1$ state can be neglected.
We therefore restrict our considerations to a two--level system, where
the important coupling parameters are the Rabi frequency $\Omega$ and
the detuning $\Delta$ of the rf--field. For weak coupling the
resonance condition is determined by the spatial dependence of
$\Delta$ within the Bose-Einstein condensate. After a transformation
into a co--rotating frame $\psi_{m_F}(t)\rightarrow
e^{-im_F\omega_{\text{rf}}t}\psi_{m_F}(t)$ followed by the standard
rotating wave--approximation, the equations for the condensate
wave function $\psi_c$ and the atom laser beam wave function $\psi_b$
are given by a set of coupled Gross--Pitaevskii--equations
\cite{Scott:97}
\begin{equation}
 \begin{array}{ll}
  i\hbar\frac{\partial}{\partial t} \psi_i=
    &-\frac{\hbar^2}{2m}{\mathbf{\nabla}}^2\psi_i
     +V_i({\mathbf{r}})\psi_i-m_F \hbar \omega_{\text{rf}} \psi_i\\ \\
    &+U(|\psi_i|^2+|\psi_j|^2)\psi_i+\hbar\Omega\psi_j\;,
 \end{array}
 \label{eq:2GPE}
\end{equation}
with $i,j=c,b$ and $U=4\pi\hbar^2a_s/m$. We have assumed that all
triplet scattering lengths have the same value $a_{ij}=a_s$
\cite{Inguscio:99}, and we will choose $a_s$ to be positive. The
external potentials are given by $V_c({\mathbf{r}})=\frac{m}{2}
\left(\omega_\perp^2 (x^2+z^2) +\omega_y^2 y^2 \right)+mgz$ and
$V_b({\mathbf{r}})=mgz$, where $g$ is the gravitational constant,
$\omega_\perp$ and $\omega_y$ are the trapping frequencies of the
cylindrically symmetric magnetic field and $m$ denotes the mass of the
atoms.

It has been found, using a separation ansatz for the spatial modes,
that the atom laser beam in the direction of gravity can be almost
perfectly described by an Airy--Function
\cite{Schenzle:99,Gerbier:01,Savage:00}. We therefore first consider
the behavior in the transversal $x$--direction as independent of the
other directions.

A naive, strictly one--dimensional treatment of the atom laser in the
horizontal $x$--direction would result in exactly two resonance points
$x=\pm\sqrt{2\hbar\omega_{rf}/m\omega_{\perp}^2}$ \cite{Wallis:97}.
This, however, is not a good approximation to the three dimensional
situation of the experiment. Since gravity leads to a displacement of
the condensate from the center of the magnetic field by an amount
$z_g=-g/\omega_\perp^2$, the resonance shell crosses the condensate
with very low curvature \cite{Esslinger:99} and is therefore better
approximated by a plane. This means that in the horizontal directions
output coupling happens along the full Thomas--Fermi distribution and
the initial beam wave function is a scaled down copy of the condensate
wave function \cite{Savage:00}.

While the atoms fall under gravity, the mean field potential they
experience from the condensate changes. To account for this in the
one--dimensional approximation one has to diminish the mean field
potential, $U_c(x,t)=U|\psi_c(x,z_t)|^2$ during the evolution
according to the free fall of the atoms, $z_t=z_g+gt^2/2$.

However, the principal physics of the horizontal mode and its
instability is most clearly demonstrated by first considering the case
with $g=0$, i.e.~taking $U_c(x,t)=U_c(x,z_g)$ to be constant in time.
The effective one--dimensional Gross--Pitaevskii equation for the beam
is then given by
\begin{equation}
 \label{eq:GP1D}
  i\hbar\frac{\partial \psi_b}{\partial t}=
      \left[-\frac{\hbar^2}{2m}\frac{\partial^2}{\partial x^2}
            +U_c(x)+U|\psi_b|^2\right]\psi_b
\end{equation}
In the Thomas-Fermi approximation the mean field potential is given by
a truncated inverted harmonic oscillator
\begin{equation}
 \label{eq:Potential}
  U_c(x)=\left\{
  \begin{array}{ll}
    \mu\left(1-\frac{x^2}{x_{TF}^2}\right)& {\text{for}}\; |x|<x_{TF},\\
    0& {\text{for}}\; |x|>x_{TF}.
  \end{array} \right.
\end{equation}
Here $\mu$ is the chemical potential and
$x_{TF}=\sqrt{2\mu/m\omega_\perp^2}$ the Thomas--Fermi radius of the
condensate.

\begin{figure}[htbp]
  \includegraphics[width=\linewidth]{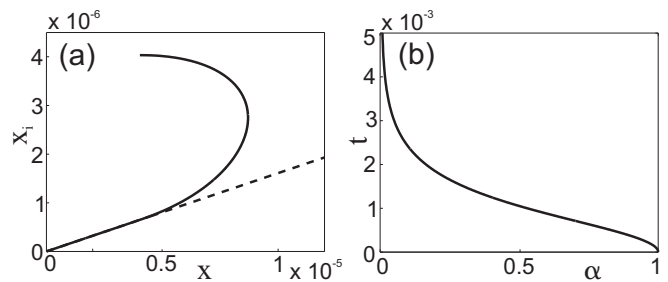}
  \caption{(a) Position of atoms starting at $x_i(t=0) < x_{TF}$ and 
    rolling off an inverted harmonic oscillator potential for a time
    of $t=2$\,ms.  The full line shows the results for a potential
    that is truncated at $x_{TF}=4\,\mu$m
    (cf.~eq.~(\ref{eq:Potential})) and the dashed line shows the
    results for an untruncated potential.  (b) Time needed for the
    atoms starting at $\alpha=x_i/x_{TF}$ to reach the condensate
    boundary at $x_{TF}$ (cf.~eq.~(\ref{eq:ttf})). The chemical
    potential in both calculations is $\mu=1700$Hz.}
  \label{fig:xf_t}
\end{figure}

Let us first analyse eq.~(\ref{eq:GP1D}) classically and neglect the
nonlinear term $U|\psi_b|^2$, because it is normally three orders of
magnitude smaller than $U_c(x)$.  Neglecting for a moment the
truncation of the potential at the condensate border, i.e., assuming
the potential ~$U_c(x)=\mu(1-x^2/x_{TF}^2),\forall x$, the classical
equation of motion
\begin{equation}
  \label{eq:NEOM}
  m\ddot x=-\frac{d U_c(x)}{d x}
\end{equation}
can be exactly integrated by $x(t)=x_i\cosh(\omega_\perp t)$, with $x_i$
the initial position of the atoms at the time of the outcoupling.
Since the $\cosh$ is an exponentially increasing function of the time,
the resulting evolution is a spreading of the initial distribution.
This means that atoms with larger $x_i$ will also have larger $x$ at
later times $t$, (see dashed line in Fig.~\ref{fig:xf_t}a).

\begin{figure}[bp]
  \includegraphics[width=\linewidth]{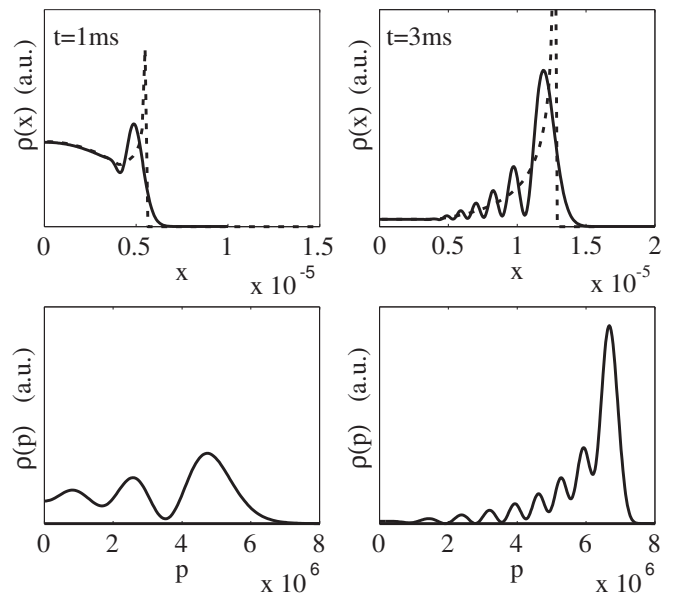}
  \caption{Transverse density distributions of the beam obtained
    from a classical (dashed) and a quantum-mechanical calculation for
    two different evolution times. The figures in the lower row show
    the absolute square of the Fourier transform of the quantum
    mechanical wave function.  A strong filamentation is clearly
    visible in the quantum results.}
  \label{fig:ClQuRollOff}
\end{figure}

The density distribution using the potential in
eq.~(\ref{eq:Potential}) is shown in Fig.~\ref{fig:ClQuRollOff}
(dashed lines).  The atoms show a temporary localization at the
Thomas--Fermi radius, since the arrival time distribution for the
atoms at the Thomas--Fermi edge
\begin{equation}
  \label{eq:ttf}
  t_{TF}(\alpha)=
    \frac{1}{\omega_{\perp}}\ln\left(\frac{1+\sqrt{1-\alpha^2}}{\alpha}\right),
\end{equation}
with $\alpha=x_i/x_{TF}$, shows a plateau (see Fig.~\ref{fig:xf_t}b).
Once the atoms have passed the Thomas--Fermi radius all their initial
potential energy, $E=\mu(1-\alpha^2)$, has been transformed into
kinetic energy and atoms starting closer to the condensate center
($x=0$) therefore end up with higher final velocity. In the asymptotic
limit, atoms originating from the center will have overtaken all other
atoms and the density distribution that originally had a negative
slope will have a positive slope. This can be seen from the dashed
curves in Fig.~\ref{fig:ClQuRollOff}.

The full quantum mechanical behaviour can be found by solving
Eqs.(\ref{eq:2GPE}) numerically, for which we use a standard
split--operator/FFT technique. As can be seen from
Fig.~\ref{fig:ClQuRollOff} (solid lines), although the general feature
of localization is preserved, the density distribution is modulated by
an interference pattern. The reason is that for finite times,
$t>t_{TF}$, atoms with different initial positions arrive at $x$ at
the same time $t$, because the equation
\begin{equation}
  \label{eq:xt}
  x(t)=x_{TF}\left(1 +\omega
  \sqrt{1-\alpha^2}\left(t-t_{TF}(\alpha)\right)\right)\;
\end{equation}
has more than one solution for $\alpha$, and quantum mechanically,
atoms coherently outcoupled at different $\alpha$'s interfere.
Observation of these fringes would prove the transversal coherence of
the atom laser beam.    

A mathematically and conceptually very elegant method to calculate the
interference pattern is by use of path integrals \cite{Feynman:65}.
The unitary wave function propagator is obtained by adding phase
factor contributions over all paths $x(t)$ along which an argument
$x_i$ in the initial wave function can evolve into the argument $x_f$
of the final state
\begin{equation}
  \label{eq:PI}
  \psi(x_f,t_f)=\int\; D[x(t)]e^{iS[x(t)]/\hbar}\psi(x_i,t_i)\;,
\end{equation}
The phase factor is the classical action along the path
\begin{equation}
 \label{eq:ClassicalAction}
  S[x(t)]=\int_{t_i}^{t_f}\;dt\;\left[\frac{1}{2}m\dot x^2-V[x(t)]\right]\;.
\end{equation}
Path integrals are usually not exactly solvable, but they can be
approximated well when for example few classical paths $x_{cl}(t)$ are
dominant. We can then restrict the integral to small deviations around
them $S[x]=S[x_{cl}+\delta x]$. This will provide a qualitative
understanding of the observed fringes as well as a very good account
of the quantitative results of the quantum calculations.

\begin{figure}[htbp]
  \includegraphics[width=\linewidth]{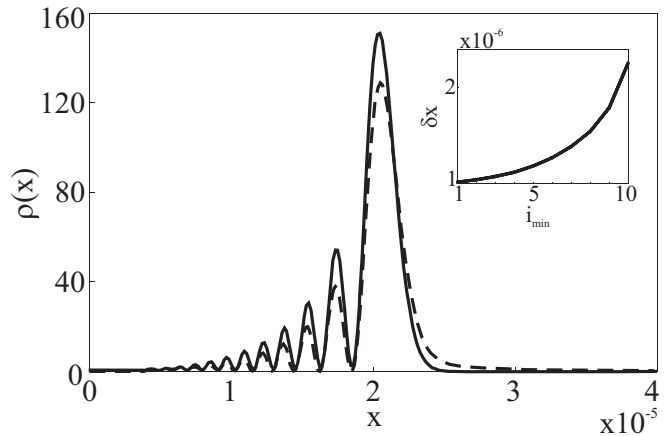}
  \caption{Transverse density distribution for t=5ms of the beam from the 
    quantum--mechanical calculation (solid line) and from the
    analytical expression based on Feynman path integrals as explained
    in the text (dashed line). The inset shows the distance $\delta x$
    between successive minima $i_{min}$ of $|\psi|^2$, starting
    at the Thomas--Fermi radius. The units of x is meters. The
    disagreement in amplitude stems from different initial conditions
    due to the different numerical methods used.}
  \label{fig:DV0G}
\end{figure}

From Fig.~\ref{fig:xf_t} one can immediately make the observation that
for $x>x_{TF}$ two classical paths contribute. The atoms emerge from
two different initial positions, and they will therefore arrive at $x$
with two different momenta $\hbar k_1$ and $\hbar k_2$.  This suggests
interference fringes of width $(|k_1-k_2|)^{-1}$, i.e.,~larger fringes
for large values of $x$ (see inset of Fig.\ref{fig:DV0G}).

%  \caption{(a) Position of atoms starting at $x_i(t=0) < x_{TF}$ and 
%    rolling off an inverted harmonic oscillator potential for a time
%    of $t=2$\,ms.  The full line shows the results for a potential
%    that is truncated at $x_{TF}=4\,\mu$m
%    (cf.~eq.~(\ref{eq:Potential})) and the dashed line shows the
%    results for an untruncated potential.  (b) Time needed for the
%    atoms starting at $\alpha=x_i/x_{TF}$ to reach the condensate
%    boundary at $x_{TF}$ (cf.~eq.~(\ref{eq:ttf})). The chemical
%    potential in both calculations is $\mu=1700$Hz.}

%\begin{figure}[htbp]
% \includegraphics[width=\linewidth,clip=true]{Fig3.eps}
%  \label{fig:DV0G}
%  \caption{Transverse density distribution for $t=5$ms of the beam from the 
%    quantum--mechanical calculation (solid line) and from the
%    analytical expression based on Feynman path integrals as explained
%    in the text (dashed line). The inset shows the distance $\delta x$
%    between successive minima $i_\text{min}$ of $|\psi|^2$, starting
%    at the Thomas--Fermi radius. The units of x is meters. The
%    disagreement in amplitude stems from different initial conditions
%    due to the different numerical methods used.}
%\end{figure}

For potentials whose second spatial derivative is constant, the
propagator can be written as \cite{Swift:82}
\begin{equation}
  \int D[x(t)]e^{\frac{i}{\hbar}S[x(t)]}=e^{\frac{i}{\hbar}S[x_{cl}]}
    \left(2\pi ik_fk_i\int_{x_i}^{x_f}
         \frac{dx}{k(x)^3}\right)^{-\frac{1}{2}}
 \label{eq:2ndOrderD}
\end{equation}
with $k_i=k(x_i)$, $k_f=k(x_f)$ and the classical action is given by
\begin{equation}
S[x_{cl}]=-E\;(t_f-t_i)+\hbar \int_{x_i}^{x_f}dx~k(x).
\end{equation}
Inserting the inverted and truncated harmonic oscillator potential of
eq.~(\ref{eq:Potential}), $V(x)=U_c(x)$, the action can be calculated
to give
\begin{align}
  S[x_{cl}]=&\frac{\mu}{\omega_{\perp}}
             \left[-\alpha^2\omega_\perp t_{TF}
             +\frac{2x_f-x_{TF}}{x_{TF}}\sqrt{1-\alpha^2}
             \right]\nonumber\\
             &-\mu(t_f-t_i)(1-\alpha^2)
\end{align}
for $x_f>x_{TF}$ and the bracketed expression in
eq.~(\ref{eq:2ndOrderD}) can be calculated to be
\begin{equation}
  \left(2\pi ik_fk_i\int_{x_i}^{x_f}
         \frac{dx}{k(x)^3}\right)^{-\frac{1}{2}}=
      \sqrt{\frac{1}{2\pi
      ia_0^2}\frac{\alpha}{\sqrt{1-\alpha^2}}},
\end{equation}
where $a_0=\sqrt{\hbar/m\omega_\perp}$ .  In Fig.~\ref{fig:DV0G} the
evolution of the wave function according to the semiclassical
approximation to eq.~(\ref{eq:PI}) is shown (dashed line). Comparison
with the full quantum mechanical evolution of eq. (\ref{eq:GP1D})
shows almost perfect agreement.

\begin{figure}[htbp]
  \includegraphics[width=\linewidth,clip=true]{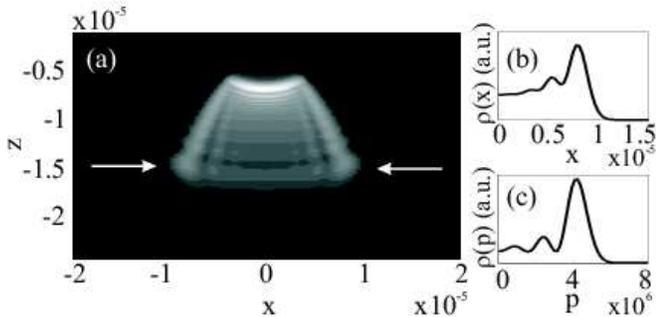}
  \caption{Two dimensional simulations of the atom laser. 
    (a) The beam after an evolution time of $2.5$\,ms.  (b) A cut
    through the density distribution at $z=-15\,\mu$m, corresponding
    to an evolution time of $1.3$\,ms. (c) The corresponding Fourier
    transform. Comparison with Fig.~\ref{fig:ClQuRollOff} shows good
    agreement. }
  \label{fig:2D}
\end{figure}

To justify and understand the semiclassical approximation, let us
note that restriction to classical paths and the assumption of
quadratic potentials is formally equivalent to a WKB approximation
\cite{Swift:82} and the condition of validity can be written as
\begin{equation}
  \frac{1}{k^2}\left|\frac{dk}{dx}\right|
  =\frac{x}{a_0^4}\frac{1}{k(x)^3}< 1
\end{equation}
where we have used the momentum of the atoms given by $\hbar
k(x)=\sqrt{2m(E-V(x))}$.  This condition is well fulfilled as long as
$V(x)$ is an inverted harmonic oscillator for all atoms.  One may,
however, question the validity close to the Thomas--Fermi radius,
where the slope of the potential changes strongly within the healing
length of the system, i.e.,~one might expect the appearance of quantum
reflection effects at this point. To estimate the reflectivity of the
potential step, we note that the de-Broglie-wavelength of the atoms
when reaching the Thomas--Fermi radius is larger than the healing
length $\xi$
\begin{equation}
 \label{eq:QR}
  \lambda_{\text{dB}}=\frac{2\pi a_0^2}{\sqrt{x_{TF}^2-x_i^2}}\geq \frac{2\pi
  a_0^2}{x_{TF}}>\frac{a_0^2}{x_{TF}}=\xi.
\end{equation}
One can therefore approximate the edge of the condensate by an
effective step. Since the exact choice of the position of the
effective potential step is not crucial, one can from
eq.~(\ref{eq:QR}) estimate its effective height to be as large as
$10\%$ of the central mean field potential. This leads to a reflection
coefficient
\begin{equation}
  \label{eq:RQ}
  R=\left|\frac{1}{1-20\sqrt{1-\alpha^2}}\right|^2 \lesssim 10^{-2}
\end{equation}
justifying very well the semiclassical treatment.

Let us finally consider a two dimensional situation including gravity.
The beam atoms are subjected to the mean-field potential for a time of
the order of $0.5-1$ms assuming a typical condensate radius of
$4\,\mu$m.  Most of the atoms therefore do not experience a complete
horizontal roll--off from the mean--field potential (compare
Fig.~\ref{fig:xf_t}b), however qualitatively the above picture remains
unchanged. We have solved the two--dimensional version of
eq.~(\ref{eq:2GPE}) in the $x$-- and $z$--plane numerically and in
Fig.~\ref{fig:2D}a the atom laser beam for short evolution times is
shown.  Once the beam has left the overlap area with the Bose-Einstein
condensate its evolution within the transverse direction is completely
determined by a free evolution and the far--field result can be simply
calculated by the Fourier transform. A cut through the density
distribution is shown in Fig.~\ref{fig:2D}b and the far field of this
distribution is shown in Fig.~\ref{fig:2D}c.  Both pictures show good
qualitative agreement with the results of the one-dimensional
analysis.

In summary we have shown that the transverse mode of an atom laser is
strongly determined by the interaction of the beam with the
mean--field of the residing condensate. Due to the finite time of this
interaction, the system resembles an interferometer in momentum space
and the beam shows filamentation in the transverse directions.

This work has been supported by the Danish Natural Science Research
Council and the Deutsche Forschungsgemeinschaft.


\begin{thebibliography}{}
\bibitem{MomentumSpread} The momentum spread in the vertical direction
  is less important, since the atoms get accelerated by gravity in
  this direction.
\end{thebibliography}
\end{document}